\documentclass[pre,superscriptaddress,showkeys,showpacs,twocolumn]{revtex4-1}
\usepackage{graphicx}
\usepackage{latexsym}
\usepackage{amsmath}
\begin {document}
\title {Phase transition and power-law coarsening in Ising-doped voter model}
\author{Adam Lipowski}
\affiliation{Faculty of Physics, Adam Mickiewicz University, Pozna\'{n}, Poland}
\author{Dorota Lipowska}
\affiliation{Faculty of Modern Languages and Literature, Adam Mickiewicz University, Pozna\'{n}, Poland}
\author{Ant\'onio  Luis Ferreira}
\affiliation{Departamento de F\'{i}sica, I3N, Universidade de Aveiro,  Portugal}
\begin {abstract}
We examine an opinion formation model, which is a mixture of Voter and Ising agents. Numerical simulations show that even a very small fraction ($\sim 1\%$) of the Ising agents drastically changes the behaviour of the Voter model.  The Voter agents act as a medium, which correlates sparsely dispersed Ising agents, and the resulting  ferromagnetic ordering persists up to a certain temperature. Upon addition of the Ising agents, a logarithmically slow coarsening of the Voter model ($d=2$), or its active steady state ($d=3$), change  into an  Ising-type power-law coarsening.
\end{abstract}

\maketitle
\section{Introduction}
Statistical mechanics models, initially intended to describe certain physical systems, find numerous applications outside the realm of physics~\cite{loreto}. The best example is probably the Ising model, which,  introduced as a model of ferromagnetism,  has been used to study collective phenomena in various nonphysical systems~\cite{stauffer}. Particularly interesting nonphysical applications of the Ising model  are related to opinion formation. This is not surprising because the dynamics of the Ising model actually tries to align neighbouring spins, similarly, perhaps, to humans deciding on their political,  religious, or consumer preferences. 
Certainly, the opinion formation is a process influenced by many factors and to take them into account, more complex models are needed. Some approaches, for example, emphasize  the role of "influentials"  for an opinion to spread~\cite{roch2005}, while some others suggest that a critical mass of individuals is needed~\cite{watts2007}. There are also models that implement the concept of social impact~\cite{holyst} or social influence~\cite{sznajd,slanina}. Other studies take into account heterogeneity of the interaction networks~\cite{barrat} as this aspect of social links seems to be  very important~\cite{wassermann}. In the aforementioned models, an opinion is represented as a discrete variable and an attempt to unify a class of such models was made by Galam~\cite{galam}.  Models with opinions   represented as continuous variables were also examined~\cite{bounded}.

For physicists, a very appealing model of opinion formation is the so-called Voter model~\cite{liggett}. The dynamics of the Voter model is very simple:  at each step a randomly selected voter takes an opinion of its randomly selected neighbour. Such tendency to align with the neighbours suggests a similarity to the Ising model, however, some subtle dynamical differences result in quite different dynamics of these models.
For our purposes, we only need to mention that the Ising model for both dimensions, $d=2$ and $d=3$, undergoes a finite-temperature transition between ferromagnetic and paramagnetic phases~\cite{huang}. Moreover, its low-temperature coarsening is curvature-driven, which is a consequence of a positive surface tension~\cite{bray}.
Actually, the curvature-driven dynamics is not restricted to  Ising (-like) models.  For example, the dynamics of certain opinion-formation models generates an effective surface tension,  as a result  of which such models share some dynamical similarities with the Ising model~\cite{baron}. With this respect, the Voter model turns out to be different and its dynamics is known to be tensionless~\cite{chate}.  Consequently, its coarsening dynamics for $d=2$ is much slower than in the Ising models, while for $d=3$ the Voter model does not coarsen at all. It should be  emphasized that such a behaviour of the Voter model is known from its exact solution~\cite{krapivsky}.  

Taking into account the heterogeneity of a human population and the multiplicity of factors affecting opinion-formation processes, a homogeneous model, wherein each agent acts in accordance with the  same rules, must certainly be unrealistic. The qualitatively different dynamics of the Ising and the Voter models prompted us to examine a model being a mixture of them.
Naively, one might expect that with this strategy, it is the largest fraction that determines the behaviour of the model, however, in the present paper we show that this is not always the case.

\section{Model}
In our model, we consider a $d$-dimensional Cartesian lattice of linear size $L$ with periodic boundary conditions. On each site~$i$ of the lattice, there is an agent, represented as a binary variable $s_i=\pm 1$, which evolves according to the Ising- or Voter-model dynamics. Initially, each agent is assigned the type of dynamics, to which it is subject: with probability $p$ the agent is set to operate according to the heat-bath Ising dynamics, and with probability $1-p$ according to the Voter dynamics. Our model is thus a quenched mixture of the Ising and Voter variables. An elementary step of the dynamics is defined as follows. 
\begin{itemize}
\item  Select an agent, say $i$. 
\item If the variable $s_i$ is of the Ising type, update it according to the heat-bath dynamics, namely, set as +1 with probability 
\begin{equation}
r(s_i\!=\!1)=\frac{1}{1+\exp(-2h_i/T)}, \ \  h_i=\sum_{j_i} s_{j_i},
\label{heat-bath}
\end{equation}
and  as $-1$  with probability $1 - r(s_i\!\!=\!\!1)$.
The temperature-like parameter~$T$ controls the noise of the system and the summation
in Eq.~(\ref{heat-bath}) is over the nearest neighbours of the agent~$i$. 
\item If the variable $s_i$ is of the Voter type,  select one of its nearest neighbours, say $j$, and set $s_i=s_j$. 
\end{itemize}

For the purpose of dynamical simulations, we define a unit of time $(t=1)$ as $N$ elementary steps,  
where $N=L^d$ is the number of sites.

Let us notice that  models combining two kinds of dynamics have already been examined. For example, Hurtado {\it et al.}~\cite{hurtado}, motivated by the nonequilibrium behaviour of interfaces in some disordered systems,  analysed an Ising model with spin variables evolving according to the heat-bath dynamics but with a randomly switched  temperature. In our case, however, the type of the dynamics used by a given agent is initially assigned and kept fixed. Moreover, our model combines two qualitatively different dynamics, which, as we will show,  leads to somewhat unexpected behaviour.

\section{Results}
To analyse our model, we used numerical simulations. Only for $p=1$, the model preserves the detailed-balance condition and is a dynamical realization of the equilibrium Ising model. 
For $p<1$, the relation with the Ising model is lost, nevertheless,  for convenience, we still use the terminology and quantities such as in  ordinary Ising model simulations. We measured magnetization $m=\frac{1}{N}\sum_i s_i$  and energy $E=-\frac{1}{N}\sum_{<i,j>} s_is_j$, where the summation is over pairs of the nearest neighbours. We also calculated  the high-temperature (at $m=0$) variance of magnetization $\chi=\frac{1}{N}(\sum_i s_i)^2$, which up to the temperature factor is the analogue of the Ising model susceptibility. 
Since our model incorporates certain quenched disorder, to calculate $m$, $E$, or $\chi$ we also averaged over independent samples (due to self-averaging, sample-to-sample fluctuations are not large and usually 10 samples were generated).
To have some insight into the dynamics of our model, we measured at $T=0$ the time dependence of the excess energy $\delta E=E-E_0$, where $E_0$ is the energy of a perfectly ferromagnetic configuration (ground state). Of course,  $E_0=-2$ for $d=2$, and  $E_0=-3$ for $d=3$. During coarsening, the so-called characteristic length $l$ increases as $l\sim t^{\phi}$ and, since $\delta E \sim l^{-1}$~\cite{shore}, it implies that $\delta E \sim t^{-\phi}$ .
In the Ising model driven by the heat-bath (i.e., nonconservative) dynamics,  $\phi=1/2$ is expected~\cite{bray}. 

\subsection{d=2}
\begin{figure}
\includegraphics[width=\columnwidth]{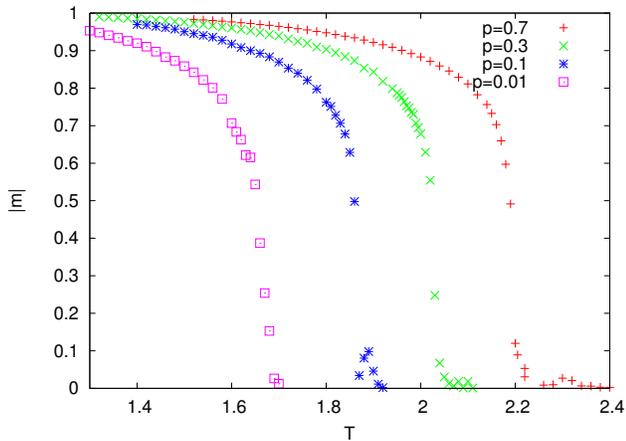}
\vspace{-8mm}
\caption{(Color online) Temperature dependence of the absolute average magnetization  $|m|$ in the two-dimensional model. Simulations were made for $L=1000$ with simulation time  $ t=5\cdot 10^5$. 
For each temperature the initial configuration was ferromagnetic ($s_i=1$).
To reach  the stationary regime, the model relaxed and the relaxation time was equal to the simulation time.
}
\label{magnet2d}
\end{figure}

\begin{figure}
\includegraphics[width=\columnwidth]{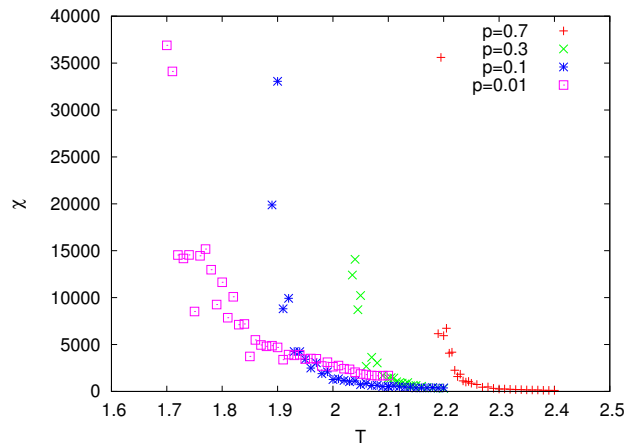}
\vspace{-8mm}
\caption{(Color online) Temperature dependence of susceptibiility  $\chi$ in the two-dimensional model. Simulations were made for $L=1000$ with simulation time (equal to relaxation time)  $ t=5\cdot 10^5$. For each temperature, the initial configuration was ferromagnetic ($s_i=1$).}
\label{susc2d}
\end{figure}

First, we describe the results of simulations for \mbox{$p=0.7$}. In such a case, the concentration of Ising agents is large and the existence of a ferromagnetic ordering at low temperature is not surprising (Fig.~\ref{magnet2d}). 
Indeed, voters (of a rather small concentration $1-p=0.3$) might be considered here as a perturbation (quenched dynamic disorder) to the pure Ising model. Since Ising agents are distributed with concentration above the percolation threshold ($p_c=0.592\dots$~\cite{newman}), an intuition based on the studies of Ising models on diluted lattices~\cite{wortis} suggests that perhaps also in this case the ferromagnetic ordering  should exist for sufficiently low temperatures. Indeed, our simulations show that the phase transition for $p=0.7$ takes place at $T\approx 2.2$, which is only slightly lower than the critical temperature in the pure Ising model ($T_c=2.269\dots$~\cite{kramers}). The peak  of the susceptibility around $T=2.2$  gives further evidence of such a transition (Fig.~\ref{susc2d}). We do not estimate the critical exponents $\beta$ and $\gamma$, describing the behaviour of $m$ and $\chi$ at the criticality, but it is, in our opinion, quite plausible that they take the pure Ising values ($\beta=1/8$ and $\gamma=1.75$). 

For lower concentrations of Ising agents, $p=0.3$, 0.1, and 0.01, basically the same behaviour can be seen (Fig.~\ref{magnet2d}, Fig.~\ref{susc2d}). Taking into account that such concentrations  are below the percolation threshold,  the existence of an apparently similar phase transition is quite surprising.  Indeed, below the percolation threshold, Ising agents form only finite (separate) clusters, which, nevertheless, get aligned  at a sufficiently low temperature, arguably due to Voter agents acting as some kind of a correlating medium. Although the surrounding Voter agents are not directly exposed to the thermal noise, at a sufficiently large temperature such a  ferromagnetic ordering is destroyed. Let us notice that this behaviour is observed even for a very small concentration of Ising agents $p=0.01$. Checking whether it extends to any  $p>0$ would require, however, further analysis and more extensive simulations.  

Our results show that at high temperature, even a tiny fraction of Ising agents  can destroy the dynamics of the Voter model: due to a destabilizing effect of Ising agents, the system does not evolve toward a single-opinion state but remains in the $m=0$ state.  In the context of opinion formation, it shows that the Voter model is actually very fragile with respect to agents operating according to the Ising dynamics.

\begin{figure}
\includegraphics[width=\columnwidth]{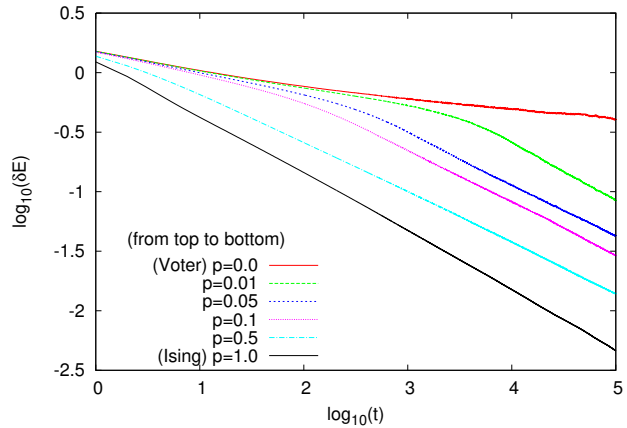}
\vspace{-8mm}
\caption{(Color online) Time dependence of the energy difference $\delta E$ for $d=2$ and several values of $p$. Simulations were made for $T=0$,  $L=2000$ and the results are averages over 100 independent samples. For $p=1$, the least square fit to our data in the range $10^2-10^5$ gives $\delta E \sim t^{-\phi}$ with $\phi=0.501(2)$.}
\label{time2d}
\end{figure}

\begin{figure}
\includegraphics[width=\columnwidth]{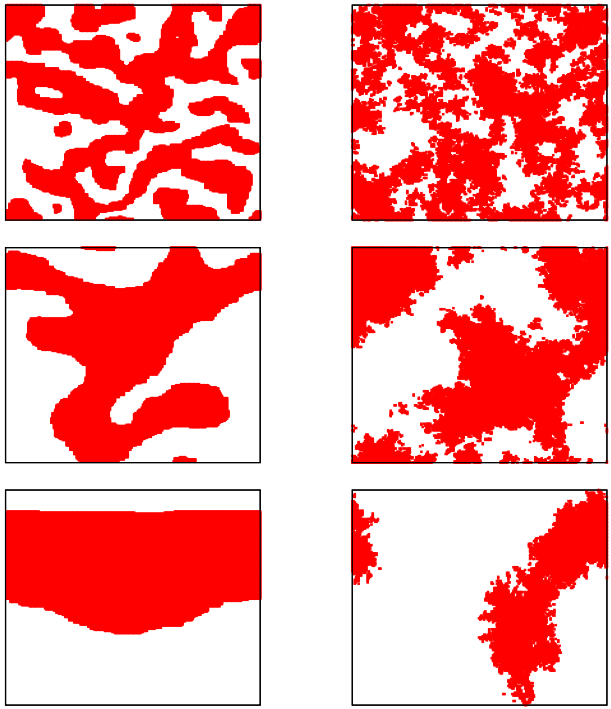}
\vspace{-8mm}
\caption{Time evolution of the $d=2$ model at $T=0$ and for $p=1$ (left) or $p=0.1$ (right) at $t=10^2$ (top), $t=10^3$ (middle), and $t=10^4$ (bottom), starting from  a random initial configuration. For the pure Ising model (left), the horizontal stripe will smooth out and the system will get stuck in a metastable configuration. Simulations  were made for $L=300$.}
\label{config}
\end{figure}

We also examined the $T=0$ dynamics of our model starting from a random initial configuration.
In particular, we calculated the time dependence of the average excess energy $\delta E=E-E_0$.
For the Ising model ($p=1$), our results (Fig.~\ref{time2d}) confirm the expected $\delta E\sim t^{-1/2}$ decay~\cite{bray}. Let us notice, however, that in such a case the system not always end up in a homogeneous ground state, but might get trapped in a certain metastable configuration. Such configurations, which  appear also during the coarsening of the $d=3$ model~\cite{lip99,olejarz}, consist of some stripes and its formation is shown in Fig.~\ref{config} (left panel). A very good agreement of our estimation of $\phi=0.501(2)$ with the expected value 1/2 shows that formation of such structures does not affect the asymptotic decay of $\delta E$.  

The Voter agents, present in our model for $p<1$, introduce the so-called interfacial noise~\cite{chate,liplip2017}. Such noise precludes formation of stripe-like configurations (Fig.~\ref{config}), but the fluctuations that it generates might slow down the dynamics. As a result,  $\phi$  slightly varies with $p$ (Fig.~\ref{time2d}). For example, for $p=0.5$, we obtain $\phi=0.433(5)$, and for smaller $p$, $\phi$ seems to increase. Even for a very small fraction of Ising agents, $p=0.01$, the fit for the data with $10^4<t<10^5$ gives $\phi=0.482(5)$, which is close to the Ising value 1/2. Only for $p=0$, when our model becomes equivalent to the Voter model,  we observe a much slower, and perhaps logaritmically slow, decay of $\delta E$, which is consistent with the exact solution~\cite{krapivsky}. Thus, the dynamical characteristics also show that the tensionless Voter model dynamics in the presence of even a very small fraction of Ising agents is replaced with the curvature-driven dynamics. Definite resolution whether for $0<p<1$ the exponent $\phi$ takes the Ising value 1/2 or is somewhat smaller (possibly varing with $p$) might require longer simulations. Let us notice, however, that $\phi=0.45$ was already reported for some other nonequilibrium models with absorbing states \cite{castello2006,dallasta2008,chate2005,vazques}.

Stability of voter dynamics and its possible split into Ising and directed percolation transition  was examined in some other models with symmetric and double degenerate absorbing states \cite{drozlip,chate2005,vazques}. Similarly to our work, these studies also  indicate fragility of voter dynamics albeit with respect to some other perturbations, e.g. longer range of interactions.

\subsection{d=3}
\begin{figure}
\includegraphics[width=\columnwidth]{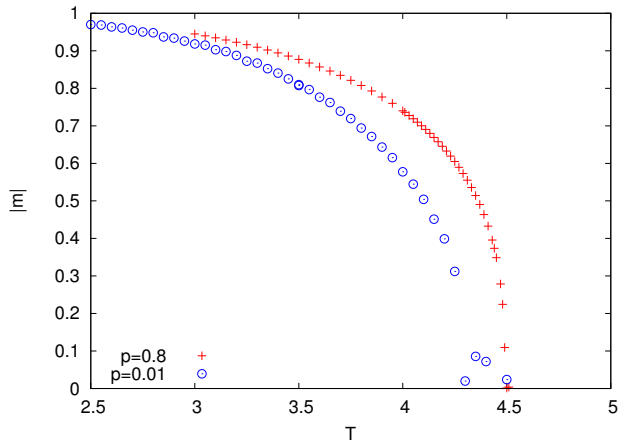}
\vspace{-8mm}
\caption{(Color online) Temperature dependence of the absolute average magnetization  $|m|$ in the three-dimensional model. Simulations were made for $L=100$.}
\label{magnetd3}
\end{figure}

We also analysed the behaviour of the $d=3$ version of our model. The steady-state magnetization behaves similarly as in the $d=2$ case. For large concentration of Ising agents ($p=0.8$), the phase transition takes place around $T=4.5$ (Fig.~\ref{magnetd3}), which is close to the $d=3$ Ising model value $4.511\dots$~\cite{binder}. However, a similar behaviour can be seen for the concentration of Ising spins $p=0.01$ with only a slightly reduced transition temperature. 

\begin{figure}
\includegraphics[width=\columnwidth]{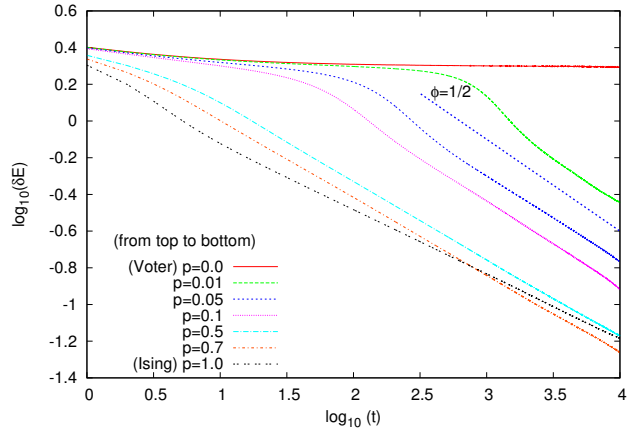}
\vspace{-8mm}
\caption{(Color online) Time dependence of the energy difference $\delta E$ for $d=3$ and several values of $p$. Simulations were made for $T=0$,  $L=300$ and the results are averages over 100 independent samples.
Dashed thick line has a slope corresponding to $\phi=1/2$.}
\label{timed3}
\end{figure}

An estimation of the exponent $\phi$  based on $T=0$ coarsening dynamics for Ising  model gives $\phi=0.358(2)$ (Fig.~\ref{timed3}), which is considerably lower than 1/2. A similar value ($\phi\sim 0.35-0.37$) was reported also in some other studies~\cite{shore} and it was suggested that a slower kinetics might be related to formation of stripe-like structures~\cite{lip99}. Actually, $T=0$ coarsening of the $d=3$ Ising model leads to the formation of even more complex and slowly evolving structures \cite{olejarz1,olejarz2} and it seems that we have only limited understanding of the process  \cite{das1,das2}.

For $p<1$, our results show that $\delta E$ decays in very good agreement with $\phi=1/2$. In our opinion, the interfacial noise generated by the Voter agents roughens interfaces (Fig.~\ref{config}), which might suppress formation of metastable stripe-like configurations and eventually  restore the $\phi=1/2$ dynamics.  Only for $p=0$,  we observe that $\delta E$ saturates at a positive value, which confirms that the Voter model for $d=3$  remains in an active phase with a mixture of opinions~\cite{krapivsky}.  

Our results show that, similarly to the $d=2$ version, as few as $1\%$ of the Ising agents drastically change the Voter model dynamics. At zero temperature, it induces the $\phi=1/2$ coarsening dynamics, although the pure Voter model remains disordered (and does not coarsen) for $d=3$. Moreover, at low temperature such  small fraction of Ising agents  is able to support a ferromagnetic ordering. Similarly to the $d=2$ case, Voter agents might be considered as correlating sparsely distributed Ising agents. However, while for $d=2$ the correlating effect of the Voter agents seems plausible (the $d=2$ Voter model exhibits a slow coarsening), for $d=3$ it is somewhat surprising beacuse in this case the Voter model remains disordered.  Overall, in the $d=3$ case, the dynamics of the Voter model is also very fragile with respect to perturbation with Ising agents.   

\section{Conclusions}
The Ising  and the Voter models are two very important models in statistical mechanics.  Although in  both models, there is a tendency for neighbouring spins/agents to align at the same value, certain dynamical differences lead to a much different behaviour. 
In the Ising model, a surface tension, present at low temperatures, induces the so-called curvature-driven dynamics with a characteristic power-law coarsening. In the Voter model, the surface tension is absent and the model either coarsens  very slowly ($d=2$) or does not coarsen at all ($d=3$).
In the present paper, we analysed a model being a mixture of these two models.
As a main result, we show that the Voter model dynamics is very unstable and even a very small ($\sim 1\%$) fraction of Ising agents changes the tensionless dynamics into the curvature-driven Ising-like dynamics with a power-law coarsening and a finite-temperature phase transition. Since the Ising agents are distributed much below the percolation threshold, it means that it is the Voter agents that correlate the Ising ones. While for $d=2$, the Voter model slowely coarsens and this correlating effect seems plausible, for $d=3$ it is more surprising. Probably, the density of the Ising agents~$p$ examined in our simulations is still sufficiently large so that a Voter medium can correlate them. Possibly,  at a lower concentration~$p$  such correlation would not take place and the system would remain disordered. Verifying such a scenario is, however, left as a future problem. 

In the context of opinion formation, our results show that in the two-dimensional case, a single-opinion state, asymptotically reached in the pure Voter model, becomes destabilized by a small fraction of the Ising agents provided that they are kept at a suficiently large temperature. Let us notice that the Ising agents might be also interpreted as voters, though such that take into account the opinion of the majority rather than that of a randomly selected neighbour. The so-called majority Voter model was extensively studied and its close relations to the Ising model are firmly established~\cite{majority1,majority2}. In our model, in the three-dimensional case, a small fraction of the Ising agents can order the Voter agents
provided that the Ising ones are kept in a sufficiently low temperature. It would be certainly interesting to examine our model on complex networks, which seem to be more suitable to describe social interactions.

In statistical mechanics, there is a lot of examples showing that the Ising dynamics emerges in a number of systems, also nonequilibrium ones, provided that their dynamics preserves an up-down symmetry~\cite{marro}. Although the Voter model also satisfies the up-down symmetry, its somewhat specific dynamics seems to be an exception. We have shown, however, that this exception is actually 
very fragile and when exposed to a small perturbation, it is replaced with a robust Ising-like dynamics. 

Acknowledgments: A.L Ferreira acknowledges support by FEDER funds through the COMPETE 2020 Programme and National Funds throught FCT—Portuguese Foundation for Science and Technology under the project UID/CTM/50025/2013.


\begin{thebibliography}{}

\bibitem{loreto} C. Castellano, S. Fortunato, and V. Loreto, Statistical physics of social dynamics, Rev. Mod. Phys. {\bf 81}, 591 (2009).

\bibitem{stauffer} D. Stauffer, Social applications of two-dimensional Ising models, Am. J. Phys. {\bf 76}, 470 (2008).

\bibitem{roch2005} C. H.  Roch, The dual roots of opinion leadership,
J. Polit. {\bf 67}, 110 (2005).

\bibitem{watts2007} D. J. Watts and P. S. Dodds, Influentials, networks, and public opinion formation,  J. Cons. Res. {\bf 34}, 441 (2007).

\bibitem{holyst} J. A. Ho\l yst,  K. Kacperski, and F. Schweitzer, Phase transitions in social impact models of opinion formation, Physica~A {\bf 285}, 199 (2000).

\bibitem{sznajd} K. Sznajd-Weron and J. Sznajd, Opinion evolution in closed community, Int. J. Mod. Phys.~C {\bf 11}, 1157 (2000).

\bibitem{slanina}  F. Slanina, H. Lavi\v cka, Analytical results for the Sznajd model of opinion formation, Eur. Phys. J.~B, {\bf 35}, 279 (2003).

\bibitem{barrat} C. Nardini, B. Kozma, and A. Barrat, Who's talking first? Consensus or lack thereof in coevolving opinion formation models, Phys. Rev. Lett. {\bf 100}, 158701 (2008).

\bibitem{wassermann} S. Wasserman and K. Faust, {\em Social Network Analysis: Methods and Applications} (Cambridge University Press, Cambridge, 1994).

\bibitem{galam} S. Galam, Local dynamics vs.\ social mechanisms: A unifying frame, EPL  {\bf 70} 705 (2005).

\bibitem{bounded} G. Deffuant, D. Neau, F. Amblard, and G. Weisbuch, Mixing beliefs among interacting agents, Adv. Complex Syst. {\bf 3}, 87 (2000).

\bibitem{liggett} T. M. Liggett, {\em Interacting Particle Systems} (Springer Verlag, New York, 1985).

\bibitem{huang} K. Huang, {\em Statistical Mechanics}  (New York, Wiley, 1987), 2nd edition.

\bibitem{bray} A. J. Bray, Theory of phase-ordering kinetics, Adv. Phys. {\bf 43}, 357 (1994).

\bibitem{baron} A. Baronchelli, L. DallAsta, A. Barrat, and V. Loreto, 
Topology-induced coarsening in language games, Phys. Rev.~E {\bf 73}, 015102 (2006).

\bibitem{chate} I. Dornic, H. Chat\'e, J. Chave, and H. Hinrichsen,   Critical coarsening without surface tension: The universality class of the voter model, Phys. Rev. Lett. {\bf 87}, 045701 (2001).

\bibitem{krapivsky} L. Frachebourg and P. L. Krapivsky, Exact results for kinetics of catalytic reactions, Phys. Rev.~E {\bf 53}, R3009 (1996).

\bibitem{hurtado} P. I. Hurtado, P. L. Garrido, and J. Marro, Analysis of the interface in a nonequilibrium two-temperature Ising model, Phys. Rev.~B {\bf 70}, 245409 (2004).

\bibitem{shore} J. D. Shore, M. Holzer, and J. P. Sethna, Logarithmically slow domain growth in nonrandomly frustrated systems: Ising models with competing interactions, Phys. Rev.~B {bf 46}, 11376 (1992).

\bibitem{newman} M. E. J.  Newman and R. M. Ziff, Efficient Monte-Carlo algorithm and high-precision results for percolation, Phys. Rev. Lett. {\bf 85 (19)}, 4104 (2000).

\bibitem{wortis} C. Jayaprakash, K. Riedel, and M. Wortis, Critical and thermodynamic properties of the randomly dilute Ising model, Phys. Rev.~B {\bf 18}, 2244 (1978).

\bibitem{kramers} H. A. Kramers and G. H. Wannier, Statistics of the two-dimensional ferromagnet, Phys. Rev. {\bf 60}, 252 (1941).

\bibitem{lip99} A. Lipowski, Anomalous phase-ordering kinetics in Ising model, Physica~A {\bf 268}, 6 (1999).

\bibitem{olejarz} V. Spirin, P. L. Krapivsky, and S. Redner, Freezing in Ising ferromagnets, Phys. Rev.~E {\bf 65}, 016119 (2001).

\bibitem{liplip2017} A. Lipowski, D. Lipowska, and A. L. Ferreira, Agreement dynamics on directed random graphs, J.~Stat. Mech. 063408 (2017).

\bibitem{castello2006} X. Castell\'o, V. M. Egu\'iluz, and M. San Miguel, Ordering dynamics with two non-excluding options: bilingualism in language competition, N. J. Phys. {\bf 8}, 308 (2006).

\bibitem{dallasta2008} L. Dall'Asta and T. Galla, Algebraic coarsening in voter models with intermediate states, J. Phys. A. {\bf 41}, 435003 (2008).

\bibitem{chate2005} O. Al Hammal, H. Chat\'e, I. Dornic, and M. A. Munoz, Langevin description of critical phenomena with two symmetric absorbing states, Phys. Rev. Lett. {\bf 94} 230601 (2005).

\bibitem{vazques} F. Vazquez, and C. L\'opez,  Systems with two symmetric absorbing states: relating the microscopic dynamics with the macroscopic behavior, Phys. Rev. E {\bf 78}, 061127 (2008).

\bibitem{drozlip} M. Droz, A. L. Ferreira, and A. Lipowski, Splitting the voter Potts model critical point, Phys. Rev. E {\bf 67}, 056108  (2003).

\bibitem{binder}  K. Binder and E. Luijten, Monte Carlo tests of renormalization-group predictions for critical phenomena in Ising models, Phys. Rep. {\bf 344}, 179 (2001).

\bibitem{olejarz1} J. Olejarz, P. L. Krapivsky, and S. Redner, Zero-temperature relaxation of three-dimensional Ising ferromagnets, Phys. Rev.~E {\bf 83}, 051104 (2011).

\bibitem{olejarz2} J. Olejarz, P. L. Krapivsky, and S. Redner, Zero-temperature freezing in the three-dimensional kinetic Ising model, Phys. Rev.~E {\bf 83}, 030104(R) (2011).

\bibitem{das1} S. K. Das, and S. Chakraborty, Kinetics of ferromagnetic ordering in 3D Ising model: how far do we understand the case of a zero temperature quench? Euro. Phys. J. Spec. Topics {\bf 226}, 765 (2017).

\bibitem{das2} S. Chakraborty, and S. K. Das, Fractality in persistence decay and domain growth during ferromagnetic ordering: Dependence upon initial correlation, Phys. Rev. E {\bf 93}, 032139 (2016).

\bibitem{majority1} Z.-X. Wu and P. Holme, Majority-vote model on hyperbolic lattices, Phys. Rev.~E {\bf 81}, 011133 (2010).
 
\bibitem{majority2} A. L. Acu\~na-Lara and F. Sastre, Critical phenomena of the majority voter model in a three-dimensional cubic lattice, Phys. Rev.~E {\bf 86}, 041123 (2012).

\bibitem{marro} J. Marro, and R. Dickman, {\em Nonequilibrium phase transitions in lattice models} (Cambridge University Press, Cambridge, 2005).

\end{thebibliography}
\end {document}